\documentclass[usenatbib,letters]{mn2e}
\usepackage{natbib,amssymb,amsmath,times,graphicx,url}

\defcitealias{pb07}{PB07}

\title[Jets from the first core]{Collimated jets from the first core}

\author[Price, Tricco \& Bate]{Daniel J. Price$^{1}$, Terrence S. Tricco$^{1}$ and Matthew R. Bate$^{1,2}$ \\
$^{1}$Monash Centre for Astrophysics (MoCA) and School of Mathematical Sciences, Monash University, Vic 3800, Australia \\
$^{2}$School of Physics, University of Exeter, Stocker Rd, Exeter EX4 4QL, UK \\
}
\date{Submitted: 22nd Dec 2011 Revised: 25th Feb 2012 Accepted: }
\pagerange{L\pageref{firstpage}--L\pageref{lastpage}} \pubyear{2011}

\begin{document}
\label{firstpage}
\bibliographystyle{mn2e}
\maketitle

\begin{abstract}
We have performed Smoothed Particle Magnetohydrodynamics (SPMHD) simulations demonstrating the production of collimated jets during collapse of 1$M_{\odot}$ molecular cloud cores to form the `first hydrostatic core' in low mass star formation. Recently a number of candidate first core objects have been observed, including L1448 IRS2E, L1451-mm and Per Bolo 58, although it is not yet clear that these are first hydrostatic cores. Recent observations of Per Bolo 58 in particular appear to show collimated, bipolar outflows which are inconsistent with previous theoretical expectations. We show that low mass first cores can indeed produce tightly collimated jets (opening angles $\lesssim 10^{\circ}$) with speeds of $\sim 2$--$7$ km/s, consistent with some of the observed candidates. We have also demonstrated, for the first time, that such phenomena can be successfully captured in SPMHD simulations.


\end{abstract}


\section{Introduction}
 Majestic, collimated jets are a defining hallmark of the the star formation process, observed during the earliest stages of star formation when the nascent young protostar remains deeply embedded in its parental molecular cloud. Magnetic fields are the thought to provide the main mechanism for launching such outflows during protostellar collapse --- via either a centrifugal `fling' \citep{bp82} or a magnetic pressure-driven `spring' \citep{lyndenbell03} --- though the details are only just beginning to be understood via numerical models of the star formation process \citep[e.g.][]{seifriedetal11}. The earliest models of protostellar collapse by \citet{larson69}, performed in spherical symmetry, showed that gravitational collapse evolves through two distinct `stall' phases: First, the gas becomes optically thick to radiation --- trapped due to dust opacity --- leading to adiabatic heating and the formation of the ``first hydrostatic core''. This phase persists for a relatively short time ($10^{3}$--$10^{4}$ years, e.g. \citealt{by95,masanagaetal98,mi00,tomidaetal10,bate11}) before the dissociation of molecular hydrogen leads to the onset of a secondary collapse phase to form the second, or stellar, core. Recent numerical models have followed the first and second collapse in three dimensions with increasing physical realism: beginning with barotropic hydrodynamics \citep{bate98}, magnetohydrodynamics (MHD) \citep{bp06}, resistive magnetohydrodynamics \citep{machidaetal06,mim08} and radiative transfer in the flux-limited diffusion approximation \citep{bate11}.
 
 Magnetohydrodynamical calculations of either the first or both stages of collapse of isolated molecular cloud cores tend to show relatively low velocity ($v\sim 2$ km/s), wide-angled outflows produced during the first core phase \citep{tomisaka02,machidaetal06,mim08,hf08,commerconetal10,buerzleetal11} with faster, high velocity ($v \gtrsim 30$ km/s) and well collimated outflows only produced during the second collapse to form the protostar \citep{machidaetal06,bp06,mim08}. The production of collimated outflows from the second collapse phase is in line with observed protostellar jets which are very well collimated and contain velocities of up to several hundred km/s \citep*[e.g.][]{rb01,ballyetal07}, consistent with the escape velocity very close to the protostellar surface. Slower and less collimated outflows produced in simulations are generally associated with the observed wide angled molecular outflows \citep{richeretal00,arceetal07,hatchelletal07b}.
 
 Recent observations have given tantalising hints of objects detected while still in the earliest, first hydrostatic core phase: L1448 IRS2E \citep{chenetal10}, L1451-mm \citep{pinedaetal11}, and Per-Bolo 58 \citep{enochetal10,dunhametal11}. While L1451-mm shows hints of a slow ($\sim$1.3--1.7 km/s for L1451-mm, see \citealt{pinedaetal11}), wide outflow consistent with simulations of the first core phase, \citet{dunhametal11} report a slow ($v \gtrsim$ 2.9 km/s) but \emph{well-collimated} outflow (opening angles of $\sim 8^\circ$) associated with Per-Bolo 58. This is in contrast to simulations which tend to show poorly collimated outflows during the first core phase.

 In this Letter, we report on simulations, performed in ideal magnetohydrodynamics, of the collapse of an idealised $1 M_{\odot}$ molecular cloud core to the first core phase. These demonstrate that highly collimated jets with velocities of up to $\sim 7$ km/s and opening angles of $\lesssim 10^{\circ}$ may indeed be produced from the first core, provided the degree of ionisation is sufficiently high. 

\section{Methods}
\label{sec:numerics}

We solve the equations of self-gravitating, ideal MHD given by
\begin{eqnarray}
\frac{d\rho}{dt} & = & -\rho \nabla\cdot {\bf v}, \label{eq:cty} \\
\frac{{\rm d}{\bf v}}{{\rm d}t} & = & -\frac{1}{\rho}\nabla\left(P + \frac12 \frac{B^{2}}{\mu_{0}} - \frac{{\bf B}{\bf B}}{\mu_{0}}\right) - \nabla\phi, \label{eq:mom} \\
\frac{\rm d}{{\rm d}t} \left(\frac{\bf B}{\rho} \right) & = & \left( \frac{{\bf B}}{\rho}\cdot \nabla \right){\bf v} \label{eq:ind}, \\
\nabla^{2}\phi & = & 4\pi G\rho, \label{eq:grav}
\end{eqnarray}
where $\rho$ is the density, ${\bf v}$ is the velocity, $P$ is the hydrodynamic pressure, ${\bf B}$ is the magnetic field, $\phi$ is the gravitational potential and $\mu_{0}$ is the permeability of free space. We solve these equations using a standard Smoothed Particle Magnetohydrodynamics (SPMHD) scheme, evolving ${\bf B}/\rho$ as the magnetic field variable (Eq.~\ref{eq:ind}), using the \citet*{bot01} source-term approach for stability, and with artificial viscosity and resistivity terms added to capture shocks and magnetic discontinuities, respectively \citep{pm05}. We use time dependent artificial viscosity and resistivity parameters as described in \citet{price12}, here using $\alpha_{\rm AV} \in [0.1,1]$ and $\alpha_{\rm B} \in [0,0.1]$. In particular, we do \emph{not} employ the Euler potentials approach, as used in previous star formation simulations \citep{pb07,pb08,pb09}, which means that there is no restriction on the geometry or winding of the field in our simulations. Instead, we control magnetic divergence errors by employing a version of the hyperbolic divergence cleaning scheme proposed by \citet{pm05}, but revised and substantially improved as described in \citet{tp12}. We find that this approach reduces the divergence errors in the collapsing core by at least two orders of magnitude, substantially reducing the non-conservation errors in momentum that previously resulted in our inability to evolve beyond the collapse phase with a ${\bf B}/\rho$-based approach \citep[see e.g.][]{pf10b,priceiaus11}. It also improves on the approach employed by \citet{buerzleetal11}, where artificial resistivity alone was used to control the divergence errors. We found artificial resistivity alone insufficient for stable and accurate long-term evolution of the jet/outflows found here.

In this paper we use a simple barotropic equation of state
\begin{equation}
P = \left\{ \begin{array}{ll}
c_{\rm s}^{2} \rho, & \rho < \rho_{\rm c} \\
c_{\rm s}^{2} \rho_{\rm c} (\rho/\rho_{\rm c})^{7/5} & \rho_{\rm c} \le \rho < \rho_{\rm d} \\
c_{\rm s}^{2}  \rho_{\rm c} (\rho_{\rm d}/\rho_{\rm c})^{7/5} \rho_{\rm d} (\rho/\rho_{\rm d})^{1.1} & \rho \ge \rho_{\rm d}\\
\end{array}\right.,
\end{equation}
where $c_{\rm s}$ is the isothermal sound speed, $\rho_{\rm c} = 10^{-14}$g/cm$^{3}$ and $\rho_{\rm d} = 10^{-10}$g/cm$^{3}$. Particles initially in the external medium are assigned a higher $c_{\rm s}$ corresponding to the higher temperature.  Sink particles \citep{bbp95} of radius 5 AU are inserted once the peak density exceeds $10^{-10}$g/cm$^{3}$, meaning that in this Letter we restrict our study to the first core phase only.

 The initial conditions are a $1 M_{\odot}$ dense, cold spherical, uniform density and slowly rotating core in pressure equilibrium with a warm, low density ambient medium. The core has radius $R_{c} = 4 \times 10^{16} $cm ($2.7 \times 10^{3}$ AU), giving an initial density of $\rho_{0} = 7.4 \times 10^{-18}$ g/cm$^{3}$ and a gravitational free-fall time of $t_{ff} = 2.4 \times 10^{4}$ yrs. We use $c_{\rm s} = 2.2\times 10^{4}$~cm~s$^{-1}$. The core is placed inside a larger, cubic domain with a side length of $8 \times 10^{16}$ cm and a $30:1$ density ratio between the core and the ambient medium, in pressure equilibrium, giving a sound speed in the external medium of $c_{\rm s,medium} = 1.2\times 10^{5}$~cm~s$^{-1}$, or $c_{\rm s,medium}^{2} \simeq 4.2 GM/R$ so that the self-gravity of the external medium is irrelevant. For simplicity we use periodic but non-self-gravitating boundary conditions on the global domain. The core is set in solid body rotation with $\Omega = 1.77 \times 10^{-13} $rad s$^{-1}$,  corresponding to a ratio of rotational to gravitational energy $\beta_{\rm r} \simeq 0.005$ and $\Omega t_{\rm ff} = 0.14$. We use $1 \times 10^{6}$ equal mass SPH particles in the core, with the density ratio giving $4.8 \times 10^{5}$ particles in the surrounding medium. We also performed calculations using $3 \times 10^{5}$ particles in the core. Resolving the Jeans length according to the \citet{bateburkert97} criterion would require $\sim 3 \times 10^{4}$ particles.
  
 The magnetic field is initially uniform in the $z$-direction, with strength $B_{0}$ characterised by the parameter $\mu$, specifying the mass-to-magnetic flux ratio ($M/\Phi$) in units of the critical value for a uniform spherical cloud \citep[e.g.][]{mestel99,mk04},
\begin{equation}
\mu \equiv \left(\frac{M}{\Phi}\right) / \left(\frac{M}{\Phi}\right)_{\rm crit},
\end{equation}
where
\begin{equation}
\left(\frac{M}{\Phi}\right) \equiv \frac{M}{\pi R_{\rm c}^{2} B_{0}}; \hspace{5mm} \left(\frac{M}{\Phi}\right)_{\rm crit} = \frac{2 c_{1}}{3} \sqrt{\frac{5}{\pi G \mu_{0} }},
\label{eq:mphicrit}
\end{equation}
where $c_{1}$ is a parameter determined numerically by \citet{ms76} to be $c_{1} \simeq 0.53$. We have performed simulations over a range of magnetic field strengths ($\mu = 20, 10, 7.5, 5, 4$ and $3$), but the main calculations we show employ $\mu = 5$, corresponding to $B_{0} = 163 \mu {\rm G}$ in physical units and an initial plasma $\beta$ (ratio of gas to magnetic pressure) of 3.3 in the core. It should be noted that the initially imposed vertical field is extremely weak compared to the 10--300 mG fields that are wound up to produce the jet.

The setup described above is otherwise identical to that employed by \citet{pb07} (hereafter \citetalias{pb07}), except that a bug later discovered in the code meant that, although \citetalias{pb07} stated that their cores were in pressure equilibrium, in fact the calculations in that paper were performed with zero pressure in the external medium. The main effect of this is that the collapse time is slightly longer and that the gas pressure at a given time is slightly lower in the collapsing core. We have here performed calculations both with and without an external confining pressure, finding similarly well-collimated jets, but at a slightly lower initial field strength when the external pressure is absent, consistent with the plasma $\beta$ being the main parameter controlling the launch/collimation of the outflows. Jets and outflows were not produced in the \citetalias{pb07} calculations because of the restrictions on the field geometry imposed by the Euler potentials formulation, meaning that the winding of the toroidal field that launches the jet (c.f. Sec.~\ref{sec:results}) was not captured.

 \begin{figure*}
    \centering
    \includegraphics[width=\textwidth]{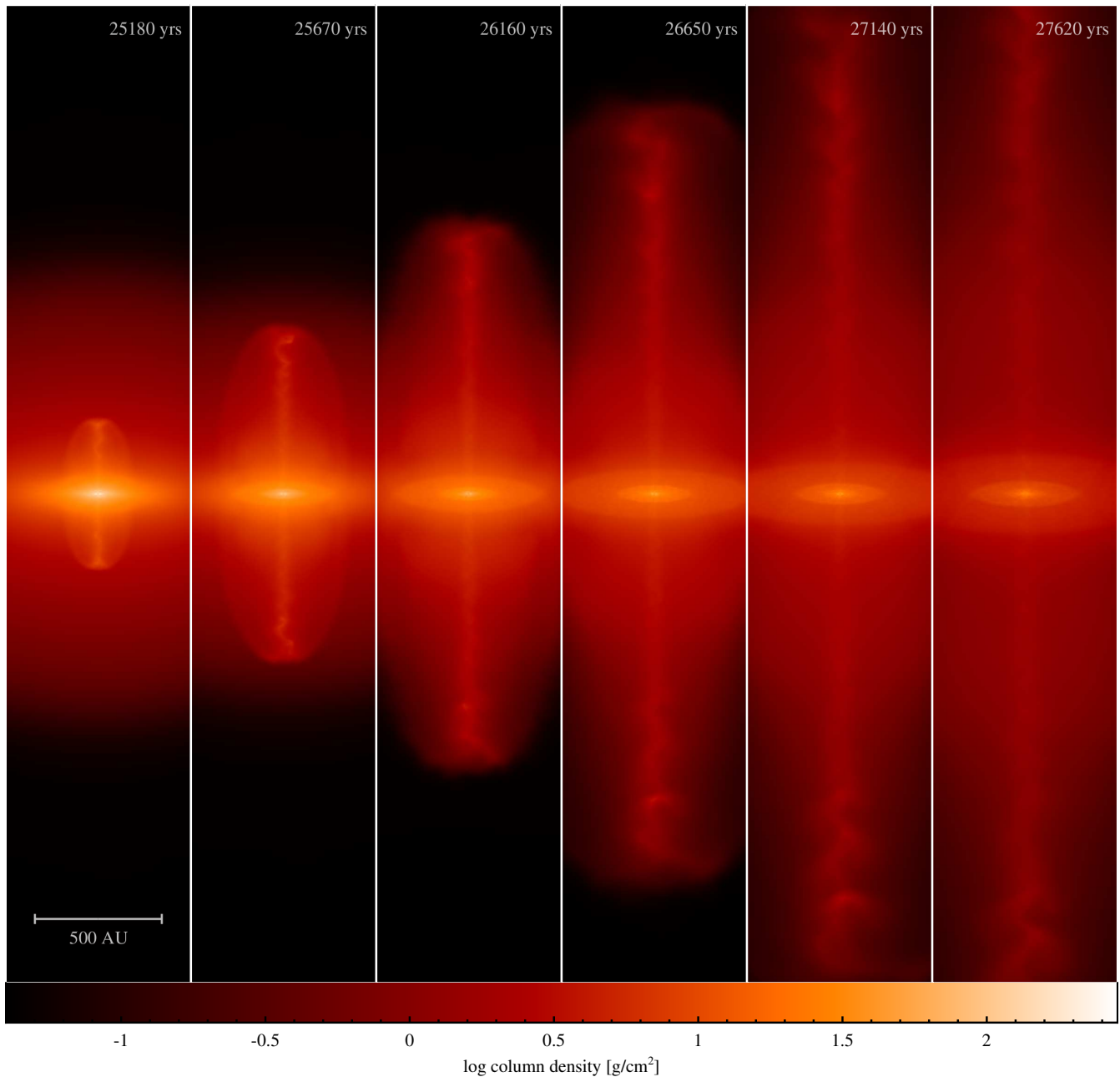} 
    \vspace{-1cm}
    \caption{Column density rendering showing the launch and propagation of the first-core jet and the entrained outflow from our simulations of a collapsing $1 M_{\odot}$ core, showing snapshots every 0.02 free-fall times (490 years). The well-collimated jet at the centre of the flow expands outwards at 3--7 km/s with an opening angle of $\lesssim 10^{\circ}$, similar to the outflow recently observed by \citet{dunhametal11} from the candidate first hydrostatic core Per Bolo 58.}
    \label{fig:jet}
 \end{figure*}

\section{Results}
\label{sec:results}
Fig.~\ref{fig:jet} shows snapshots of column density from the $\mu = 5$ calculation taken every 0.02 free-fall times (490 years) from $t/t_{\rm ff} = 1.03$--1.13, shortly after the formation of the first core. The wind up of toroidal magnetic fields in the inner disc leads to the launch of a strong, well-collimated ($\lesssim 10^{\circ}$ opening angle) jet, with outflow velocities of $\sim 5$ km/s. The jet itself can be seen to entrain a slower, wider outflow and at later times (rightmost panels) shows distinct `kinks' or `wiggles' as a result of material entrained in the helical magnetic field expanding in the z direction. At $\sim 3,000$ years after first core formation (rightmost panel) the outflow has already expanded to several thousand AU, beyond our initial core radius and well into the surrounding medium. The jet continues to be driven essentially until all of the collapsing material has been used up (this occurs at $t/t_{\rm ff} \approx 1.2$, after which material from the initially hot external medium starts to be accreted, though the outflow continues to be driven as long as mass is supplied). Although a flattened, disc-like object appears surrounding the central object, orbital velocities in the midplane are sub-Keplerian by a factor of $\sim$3--4.

\begin{figure}
    \centering
    \includegraphics[width=\columnwidth]{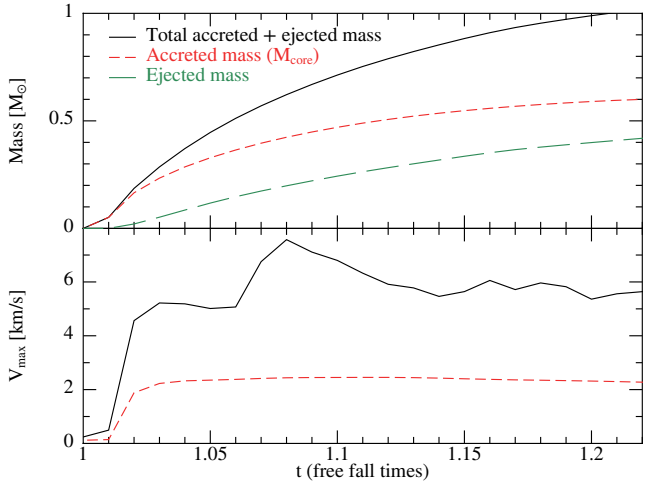} 
    \caption{Cumulative total accreted and ejected mass as a function of time (top panel, as indicated), together with the maximum velocity of ejected material (bottom panel, solid/black line) and the mean velocity of all particles with $v_{r} > 0.1$ km/s (bottom panel, red/dashed line). The outflow is remarkably efficient, expelling $40\%$ of accreted material by the end of the simulation (halted once the complete initial core mass has been accreted) at maximum speeds of 5--7 km/s (bottom panel), with a mean velocity of all outflowing material $\sim 2$ km/s.}
    \label{fig:mass}
 \end{figure}

Fig.~\ref{fig:mass} shows the accreted mass (i.e., the mass of the sink particle, red short-dashed line) and ejected mass (defined as all material with a spherical $v_{r} > 0.1$ km/s, green long-dashed line) as a function of time in the simulation, as well as the sum of these (solid line). The outflow is remarkably efficient --- with up to 40\% of the initial material in the core ejected. The lower panel shows the maximum velocity of jet material (solid line) together with the mean velocity of all particles with $v_{r} > 0.1$ km/s, indicating maximum speeds of 5--7 km/s in the jet with a mean velocity of outflowing material around 2 km/s.

\begin{figure}
    \centering
   \includegraphics[width=0.49\columnwidth]{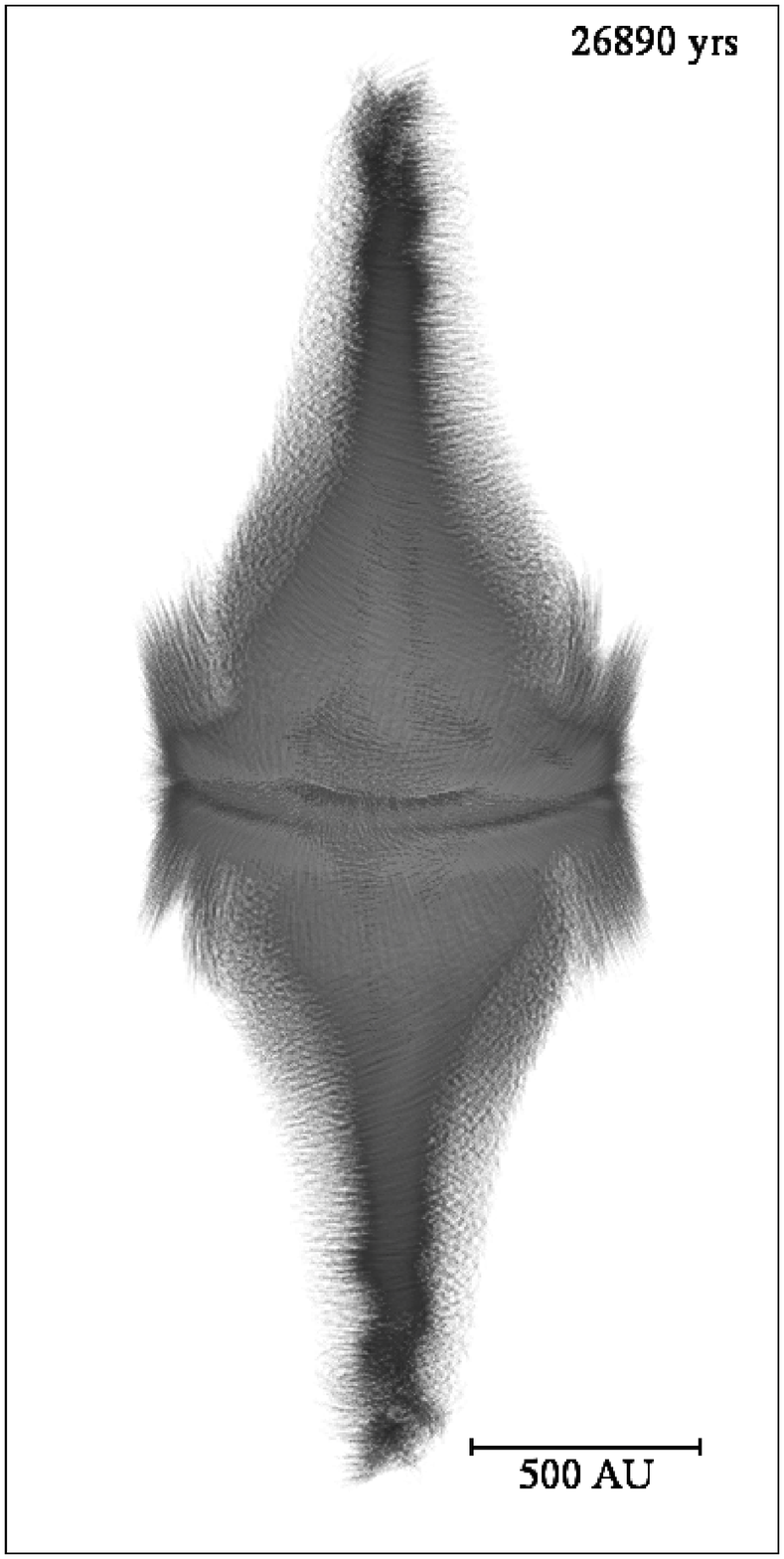} 
   \includegraphics[width=0.49\columnwidth]{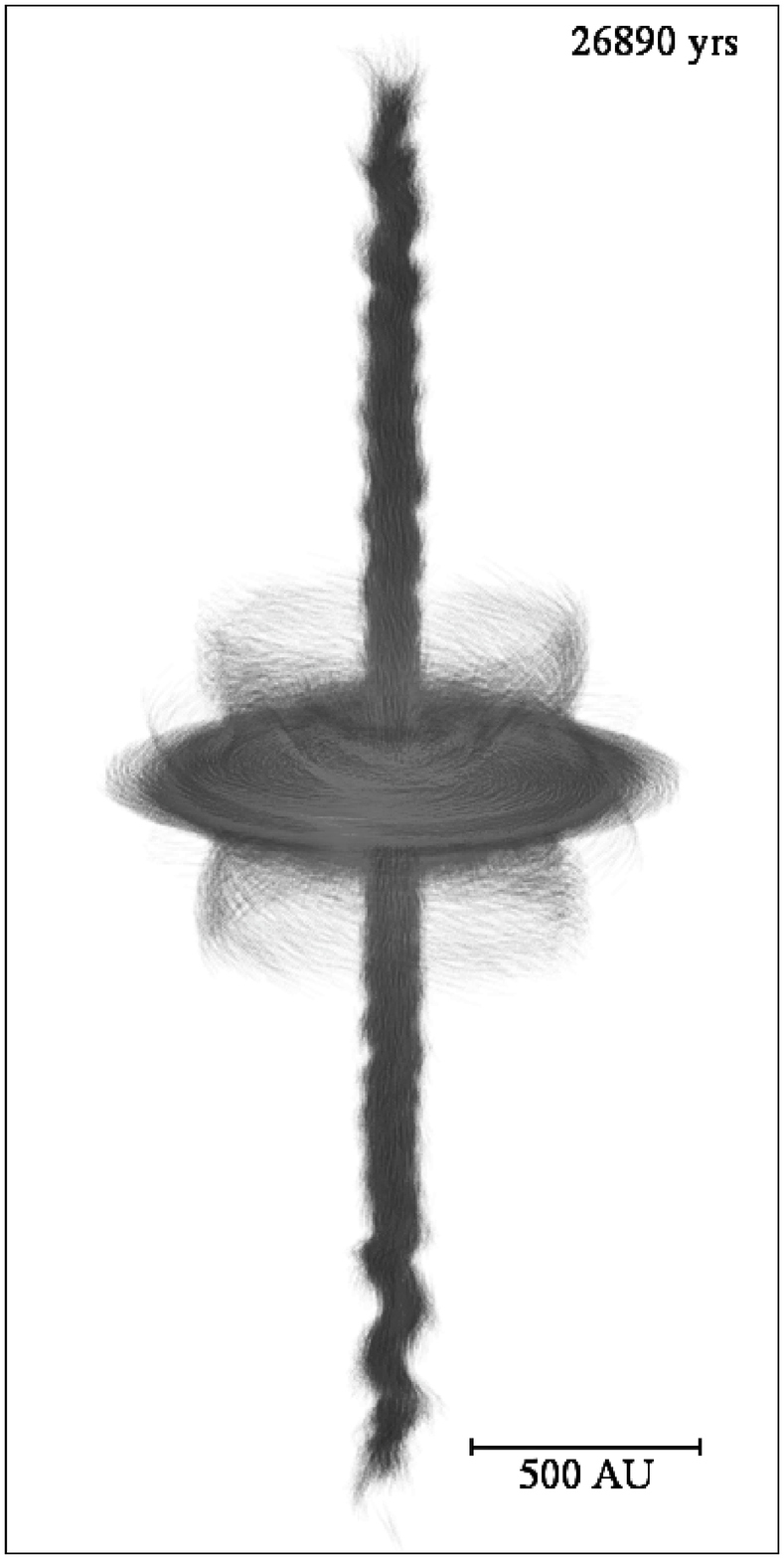} 
    \caption{Rendering of the magnetic field geometry (left) and magnetic current (right) in the $\mu = 5$ simulation, showing the strongly wound, toroidal magnetic field and strong currents that launch the jet. Wiggles and kinks seen in the column density (Fig.~\ref{fig:jet}) can also be seen in the field geometry, indicating that this material is entrained by the magnetic field. Magnetic field strengths in the jet are $\sim 10$--$300$ mG.}
    \label{fig:bfield}
 \end{figure}

Fig.~\ref{fig:bfield} shows a rendering of the magnetic field in the simulation at $t/t_{ff}=1.1$ (showing the field on each SPMHD particle drawn with an opacity proportional to the field strength). The field near the protostar is strong ($\sim 100$ mG) and tightly wound in a toroidal geometry, which is responsible for the high degree of collimation. Animations of the field evolution show clearly the growth of this ``magnetic tower'' flow \citep{lyndenbell03}.

\begin{figure}
    \centering
    \includegraphics[width=\columnwidth]{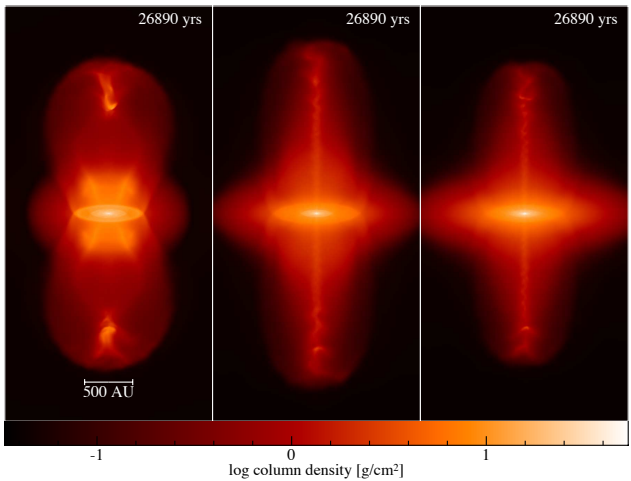} 
    \caption{Effect of varying the initial magnetic field strength. With weaker fields ($\mu = 10$, left panel) an initial fast jet is launched but detaches, leaving a transitory `knot' in a wide angled, poorly collimated wind. At higher field strengths the jet is more tightly collimated ($\mu = 5$ and $\mu = 4$, middle and right panels, respectively).}
    \label{fig:magstrength}
 \end{figure}

We find similar jets are produced for a range of magnetic field strengths, specifically for $4 \lesssim \mu \lesssim 10$ in our setup (Fig.~\ref{fig:magstrength}), corresponding to initial field strengths of $B_{0} \approx 80$--$200 \mu $G and initial plasma $\beta$'s in the range $2$--$13$. The $\mu = 10$ calculation produces a transient jet but then only a wide angled, poorly collimated wind, while simulations with $\mu \gtrsim 4$ tend to induce strong magnetic braking of the central object. Weaker field ($\mu \lesssim 20$) calculations are complicated by the fact that secondary fragmentation tends to occur in the present setup when a barotropic equation of state is used (c.f. \citealt{buerzleetal11}).

\section{Discussion}
\label{sec:discussion}
 In this Letter we have demonstrated the production of highly collimated jets with velocities of up to 7 km/s during the first hydrostatic core phase of star formation. The features are similar to the collimated jets found in other simulations, either during the second collapse phase \citep{tomisaka02,bp06,mim08}, at much higher masses \citep{seifriedetal11} or at low mass but with lower velocities \citep{ch10}.
 
  Firstly, our simulations serve to demonstrate, for the first time, the production of stable, well collimated jets with a Smoothed Particle Magnetohydrodynamics (SPMHD) technique. \citet{buerzleetal11} demonstrated that a standard artificial-resistivity based SPMHD approach \citep{pm05} could be used to simulate protostellar outflows, without the restrictions on magnetic field geometry associated with earlier Euler-potentials-based approaches \citep[e.g.][]{pb07,pb08}. We find that using artificial resistivity alone to control the divergence error is insufficient for long term, stable evolution of the jets found here and that a high resistivity is required simply to control the divergence error. Our implementation of a new divergence cleaning technique \citep[described in detail in][]{tp12} means that we have been able to follow the stable growth of the jet to several thousand AU --- until the entire molecular cloud core has been accreted --- without requiring a high resistivity, which we found could suppress the formation of the collimated flow.
  
 The jets we find are similar to those produced in other calculations, both protostellar and in accretion discs more generally \citep[see review by][]{pudritzetal07}, driven by a `magnetic tower' of tightly wound toroidal magnetic field expanding at a velocity related to the orbital velocity at the footpoint of the tower \citep{lyndenbell03}. Wiggles similar to those seen here (e.g. Fig.~\ref{fig:jet}) and a corkscrew-like morphology are a common feature arising from 3D jet simulations \citep{ouyedetal03}.

  We find the main requirements for the production of collimated outflows during the first core to be weak, but not too weak, initial magnetic fields (specifically, $4 \lesssim \mu \lesssim 10$ for the setup described here), rotational motion, and relatively low resistivity.  \citet{machidaetal06} also note that the jets produced in their (second core) simulations are sensitive to resistivity, noting that jets always occur in their calculations employing ideal MHD, but not always in their resistive MHD calculations. \citet{ch10} find similar, collimated outflows from the first core in their ideal MHD simulations, albeit with lower outflow speeds ($v \approx 1.2$--$1.8$ km/s) and dependent on the angle between the magnetic field and the rotation axis.

  


Considering the high physical resistivity thought to be present in molecular clouds at these densities \citep[e.g.][]{nnu02}, the use of ideal MHD, both here and in \citet{ch10}, is extremely \emph{un}realistic, and one would not expect jets to be produced from the first core on this basis. However, the \citet{dunhametal11} observations do suggest a collimated, bipolar outflow from a first core candidate. One possibility is that first-core jets are still possible at high resistivity ($\eta \gtrsim 10^{20}$ cm$^{2}$/s) but require higher magnetic field strengths in the initial conditions or only occur at higher density. Preliminary investigations in this regard suggest that this may indeed be the case. The other possibility is that the conductivity in Per-Bolo 58 is indeed higher than the estimates given by \citet{nnu02}, perhaps due to thermal ionisation following first core formation. Further simulations should be able to distinguish between these two possibilities. Measurements of the magnetic field strength and/or geometry in Per-Bolo 58 would also provide a critical constraint on simulation models.

  Our simulations suggest that objects such as Per-Bolo 58, showing a collimated, bipolar outflow with a characteristic velocity of $\sim 3$ km/s, may indeed be viable first core candidates. The outflow in L1451-mm is less well collimated, but remarkably efficient --- \citet{pinedaetal11} note that the amount of mass needed to keep material at 560 AU with a velocity of 1.3 km/s is $\approx 0.53$ M$_{\odot}$, almost twice the mass observed in the dense core itself. This is consistent with the high efficiency found here (Fig.~\ref{fig:mass}, see also \citealt{ch10}), where up to $40\%$ of the mass in our original core is ejected (and more if mass is continually supplied from an external reservoir). On the other hand, speeds of $\sim 25$ km/s reported for L1448 IRS2E \citep{chenetal10} would seem inconsistent with an object in the first core phase, though it should be noted that velocities $\gtrsim 9$ km/s would not be reached in the simulations shown here due to the 5 AU size of the central sink particle.

\section*{Acknowledgments}
DJP acknowledges funding via an Australian Research Council Discovery Grant (DP1094585). MRB acknowledges that: this work, conducted as part of the award ``The formation of stars and planets: Radiation hydrodynamical and magnetohydrodynamical simulations"  made under the European Heads of Research Councils and European Science Foundation EURYI (European Young Investigator) Awards scheme, was supported by funds from the Participating Organisations of EURYI and the EC Sixth Framework Programme. Calculations were performed on the University of Exeter supercomputer and the Monash Sun Grid, with thanks to Philip Chan. Visualisations were made using \textsc{splash} \citep{splashpaper}.
\bibliography{sph,mhd,starformation,jets,resistivity}

\begin{thebibliography}{}

\bibitem[\protect\citeauthoryear{{Arce}, {Shepherd}, {Gueth}, {Lee},
  {Bachiller}, {Rosen} \& {Beuther}}{{Arce} et~al.}{2007}]{arceetal07}
{Arce} H.~G.,  {Shepherd} D.,  {Gueth} F.,  {Lee} C.-F.,  {Bachiller} R.,
  {Rosen} A.,    {Beuther} H.,  2007, Protostars and Planets V, pp 245--260

\bibitem[\protect\citeauthoryear{{Bally}, {Reipurth} \& {Davis}}{{Bally}
  et~al.}{2007}]{ballyetal07}
{Bally} J.,  {Reipurth} B.,    {Davis} C.~J.,  2007, Protostars and Planets V,
  pp 215--230

\bibitem[\protect\citeauthoryear{{Banerjee} \& {Pudritz}}{{Banerjee} \&
  {Pudritz}}{2006}]{bp06}
{Banerjee} R.,  {Pudritz} R.~E.,  2006, ApJ, 641, 949

\bibitem[\protect\citeauthoryear{{Bate}}{{Bate}}{1998}]{bate98}
{Bate} M.~R.,  1998, ApJL, 508, L95

\bibitem[\protect\citeauthoryear{{Bate}}{{Bate}}{2011}]{bate11}
{Bate} M.~R.,  2011, MNRAS, 417, 2036

\bibitem[\protect\citeauthoryear{{Bate}, {Bonnell} \& {Price}}{{Bate}
  et~al.}{1995}]{bbp95}
{Bate} M.~R.,  {Bonnell} I.~A.,    {Price} N.~M.,  1995, MNRAS, 277, 362

\bibitem[\protect\citeauthoryear{{Bate} \& {Burkert}}{{Bate} \&
  {Burkert}}{1997}]{bateburkert97}
{Bate} M.~R.,  {Burkert} A.,  1997, MNRAS, 288, 1060

\bibitem[\protect\citeauthoryear{{Blandford} \& {Payne}}{{Blandford} \&
  {Payne}}{1982}]{bp82}
{Blandford} R.~D.,  {Payne} D.~G.,  1982, MNRAS, 199, 883

\bibitem[\protect\citeauthoryear{{B{\o}rve}, {Omang} \& {Trulsen}}{{B{\o}rve}
  et~al.}{2001}]{bot01}
{B{\o}rve} S.,  {Omang} M.,    {Trulsen} J.,  2001, ApJ, 561, 82

\bibitem[\protect\citeauthoryear{{Boss} \& {Yorke}}{{Boss} \&
  {Yorke}}{1995}]{by95}
{Boss} A.~P.,  {Yorke} H.~W.,  1995, ApJL, 439, L55

\bibitem[\protect\citeauthoryear{{B{\"u}rzle}, {Clark}, {Stasyszyn}, {Dolag} \&
  {Klessen}}{{B{\"u}rzle} et~al.}{2011}]{buerzleetal11}
{B{\"u}rzle} F.,  {Clark} P.~C.,  {Stasyszyn} F.,  {Dolag} K.,    {Klessen}
  R.~S.,  2011, MNRAS, 417, L61

\bibitem[\protect\citeauthoryear{{Chen}, {Arce}, {Zhang}, {Bourke},
  {Launhardt}, {Schmalzl} \& {Henning}}{{Chen} et~al.}{2010}]{chenetal10}
{Chen} X.,  {Arce} H.~G.,  {Zhang} Q.,  {Bourke} T.~L.,  {Launhardt} R.,
  {Schmalzl} M.,    {Henning} T.,  2010, ApJ, 715, 1344

\bibitem[\protect\citeauthoryear{{Ciardi} \& {Hennebelle}}{{Ciardi} \&
  {Hennebelle}}{2010}]{ch10}
{Ciardi} A.,  {Hennebelle} P.,  2010, MNRAS, 409, L39

\bibitem[\protect\citeauthoryear{{Commer{\c c}on}, {Hennebelle}, {Audit},
  {Chabrier} \& {Teyssier}}{{Commer{\c c}on} et~al.}{2010}]{commerconetal10}
{Commer{\c c}on} B.,  {Hennebelle} P.,  {Audit} E.,  {Chabrier} G.,
  {Teyssier} R.,  2010, A\&A, 510, L3

\bibitem[\protect\citeauthoryear{{Dunham}, {Chen}, {Arce}, {Bourke}, {Schnee}
  \& {Enoch}}{{Dunham} et~al.}{2011}]{dunhametal11}
{Dunham} M.~M.,  {Chen} X.,  {Arce} H.~G.,  {Bourke} T.~L.,  {Schnee} S.,
  {Enoch} M.~L.,  2011, ApJ, 742, 1

\bibitem[\protect\citeauthoryear{{Enoch}, {Lee}, {Harvey}, {Dunham} \&
  {Schnee}}{{Enoch} et~al.}{2010}]{enochetal10}
{Enoch} M.~L.,  {Lee} J.-E.,  {Harvey} P.,  {Dunham} M.~M.,    {Schnee} S.,
  2010, ApJL, 722, L33

\bibitem[\protect\citeauthoryear{{Hatchell}, {Fuller} \& {Richer}}{{Hatchell}
  et~al.}{2007}]{hatchelletal07b}
{Hatchell} J.,  {Fuller} G.~A.,    {Richer} J.~S.,  2007, A\&A, 472, 187

\bibitem[\protect\citeauthoryear{{Hennebelle} \& {Fromang}}{{Hennebelle} \&
  {Fromang}}{2008}]{hf08}
{Hennebelle} P.,  {Fromang} S.,  2008, A\&A, 477, 9

\bibitem[\protect\citeauthoryear{{Larson}}{{Larson}}{1969}]{larson69}
{Larson} R.~B.,  1969, MNRAS, 145, 271

\bibitem[\protect\citeauthoryear{{Lynden-Bell}}{{Lynden-Bell}}{2003}]{lyndenbe%
ll03}
{Lynden-Bell} D.,  2003, MNRAS, 341, 1360

\bibitem[\protect\citeauthoryear{{Mac Low} \& {Klessen}}{{Mac Low} \&
  {Klessen}}{2004}]{mk04}
{Mac Low} M.,  {Klessen} R.~S.,  2004, Rev. Mod. Phys., 76, 125

\bibitem[\protect\citeauthoryear{{Machida}, {Inutsuka} \&
  {Matsumoto}}{{Machida} et~al.}{2006}]{machidaetal06}
{Machida} M.~N.,  {Inutsuka} S.,    {Matsumoto} T.,  2006, ApJL, 647, L151

\bibitem[\protect\citeauthoryear{{Machida}, {Inutsuka} \&
  {Matsumoto}}{{Machida} et~al.}{2008}]{mim08}
{Machida} M.~N.,  {Inutsuka} S.-i.,    {Matsumoto} T.,  2008, ApJ, 676, 1088

\bibitem[\protect\citeauthoryear{{Masunaga} \& {Inutsuka}}{{Masunaga} \&
  {Inutsuka}}{2000}]{mi00}
{Masunaga} H.,  {Inutsuka} S.-I.,  2000, ApJ, 531, 350

\bibitem[\protect\citeauthoryear{{Masunaga}, {Miyama} \& {Inutsuka}}{{Masunaga}
  et~al.}{1998}]{masanagaetal98}
{Masunaga} H.,  {Miyama} S.~M.,    {Inutsuka} S.-I.,  1998, ApJ, 495, 346

\bibitem[\protect\citeauthoryear{{Mestel}}{{Mestel}}{1999}]{mestel99}
{Mestel} L.,  1999, {Stellar magnetism}.
Oxford: Clarendon

\bibitem[\protect\citeauthoryear{{Mouschovias} \& {Spitzer} Jr.}{{Mouschovias}
  \& {Spitzer}}{1976}]{ms76}
{Mouschovias} T.~C.,  {Spitzer} Jr. L.,  1976, ApJ, 210, 326

\bibitem[\protect\citeauthoryear{{Nakano}, {Nishi} \& {Umebayashi}}{{Nakano}
  et~al.}{2002}]{nnu02}
{Nakano} T.,  {Nishi} R.,    {Umebayashi} T.,  2002, ApJ, 573, 199

\bibitem[\protect\citeauthoryear{{Ouyed}, {Clarke} \& {Pudritz}}{{Ouyed}
  et~al.}{2003}]{ouyedetal03}
{Ouyed} R.,  {Clarke} D.~A.,    {Pudritz} R.~E.,  2003, ApJ, 582, 292

\bibitem[\protect\citeauthoryear{{Pineda}, {Arce}, {Schnee}, {Goodman},
  {Bourke}, {Foster}, {Robitaille}, {Tanner}, {Kauffmann}, {Tafalla}, {Caselli}
  \& {Anglada}}{{Pineda} et~al.}{2011}]{pinedaetal11}
{Pineda} J.~E.,  {Arce} H.~G.,  {Schnee} S.,  {Goodman} A.~A.,  {Bourke} T.,
  {Foster} J.~B.,  {Robitaille} T.,  {Tanner} J.,  {Kauffmann} J.,  {Tafalla}
  M.,  {Caselli} P.,    {Anglada} G.,  2011, arXiv:1109.1207

\bibitem[\protect\citeauthoryear{{Price}}{{Price}}{2007}]{splashpaper}
{Price} D.~J.,  2007, Publ. Astron. Soc. Aust., 24, 159

\bibitem[\protect\citeauthoryear{{Price}}{{Price}}{2011}]{priceiaus11}
{Price} D.~J.,  2011, in {J.~Alves, B.~G.~Elmegreen, J.~M.~Girart, \&
  V.~Trimble} ed., Computational Star Formation Vol.~270 of IAU Symposium, pp
  169--177

\bibitem[\protect\citeauthoryear{{Price}}{{Price}}{2012}]{price12}
{Price} D.~J.,  2012, J. Comp. Phys., 231, 759

\bibitem[\protect\citeauthoryear{{Price} \& {Bate}}{{Price} \&
  {Bate}}{2007}]{pb07}
{Price} D.~J.,  {Bate} M.~R.,  2007, MNRAS, 377, 77

\bibitem[\protect\citeauthoryear{{Price} \& {Bate}}{{Price} \&
  {Bate}}{2008}]{pb08}
{Price} D.~J.,  {Bate} M.~R.,  2008, MNRAS, 385, 1820

\bibitem[\protect\citeauthoryear{{Price} \& {Bate}}{{Price} \&
  {Bate}}{2009}]{pb09}
{Price} D.~J.,  {Bate} M.~R.,  2009, MNRAS, 398, 33

\bibitem[\protect\citeauthoryear{{Price} \& {Federrath}}{{Price} \&
  {Federrath}}{2010}]{pf10b}
{Price} D.~J.,  {Federrath} C.,  2010, in {N.~V.~Pogorelov, E.~Audit, \&
  G.~P.~Zank} ed., Numerical Modeling of Space Plasma Flows, Astronum-2009
  Vol.~429 of ASP Conf. Ser., p.~274

\bibitem[\protect\citeauthoryear{{Price} \& {Monaghan}}{{Price} \&
  {Monaghan}}{2005}]{pm05}
{Price} D.~J.,  {Monaghan} J.~J.,  2005, MNRAS, 364, 384

\bibitem[\protect\citeauthoryear{{Pudritz}, {Ouyed}, {Fendt} \&
  {Brandenburg}}{{Pudritz} et~al.}{2007}]{pudritzetal07}
{Pudritz} R.~E.,  {Ouyed} R.,  {Fendt} C.,    {Brandenburg} A.,  2007,
  Protostars and Planets V, pp 277--294

\bibitem[\protect\citeauthoryear{{Reipurth} \& {Bally}}{{Reipurth} \&
  {Bally}}{2001}]{rb01}
{Reipurth} B.,  {Bally} J.,  2001, Ann. Rev. Astron. Astroph., 39, 403

\bibitem[\protect\citeauthoryear{{Richer}, {Shepherd}, {Cabrit}, {Bachiller} \&
  {Churchwell}}{{Richer} et~al.}{2000}]{richeretal00}
{Richer} J.~S.,  {Shepherd} D.~S.,  {Cabrit} S.,  {Bachiller} R.,
  {Churchwell} E.,  2000, Protostars and Planets IV, p.~867

\bibitem[\protect\citeauthoryear{{Seifried}, {Pudritz}, {Banerjee}, {Duffin} \&
  {Klessen}}{{Seifried} et~al.}{2011}]{seifriedetal11}
{Seifried} D.,  {Pudritz} R.~E.,  {Banerjee} R.,  {Duffin} D.,    {Klessen}
  R.~S.,  2011, arXiv:1109.4379

\bibitem[\protect\citeauthoryear{{Tomida}, {Machida}, {Saigo}, {Tomisaka} \&
  {Matsumoto}}{{Tomida} et~al.}{2010}]{tomidaetal10}
{Tomida} K.,  {Machida} M.~N.,  {Saigo} K.,  {Tomisaka} K.,    {Matsumoto} T.,
  2010, ApJL, 725, L239

\bibitem[\protect\citeauthoryear{{Tomisaka}}{{Tomisaka}}{2002}]{tomisaka02}
{Tomisaka} K.,  2002, ApJ, 575, 306

\bibitem[\protect\citeauthoryear{{Tricco} \& {Price}}{{Tricco} \&
  {Price}}{2012}]{tp12}
{Tricco} T.~S.,  {Price} D.~J.,  2012, submitted to J. Comp. Phys.

\end{thebibliography}

\label{lastpage}
\enddocument